\pdfoutput=1
\documentclass[12pt,letterpaper]{article}
\usepackage{jheppub}
\usepackage{ulem}

\newcommand{\e}{\epsilon}
\renewcommand{\l}{\ell}
\renewcommand{\O}{\mathcal{O}}
\newcommand{\smm}{{\rm\rule[2.4pt]{6pt}{0.65pt}}}
\newcommand{\smp}{\hspace{0.5pt}\text{{\small+}}\hspace{-0.5pt}}
\newcommand{\ab}[1]{\langle #1\rangle}
\newcommand{\figBox}[4]{\mbox{\hspace{#1 cm}\raisebox{#2 cm}{\includegraphics[scale=#3]{#4}}}}
\newcommand{\newcap}{{\small\mathrm{\raisebox{0.95pt}{{$\,\bigcap\,$}}}}}

\title{Collinear and Soft Limits of Multi-Loop Integrands
in $\mathcal{N}=4$ Yang-Mills}

\author{John Golden,}
\author{Marcus Spradlin}

\affiliation{Department of Physics, Brown University, Box 1843,
Providence, RI 02912 USA}

\abstract{
It has been argued in arXiv:1112.6432 that the planar four-point integrand
in $\mathcal{N}=4$ super Yang-Mills theory is uniquely determined by
dual conformal invariance together with the absence of a double
pole in the integrand of the logarithm in the limit
as a loop integration variable
becomes collinear with an external momentum.  In this paper we reformulate
this condition in a simple way in terms of the amplitude itself,
rather than its logarithm,
and verify that it holds for two- and three-loop MHV
integrands for $n>4$.
We investigate the extent to which this collinear constraint and a constraint
on the soft behavior of integrands can be used to determine integrands.
We find an interesting complementarity whereby the soft constraint becomes
stronger while the collinear constraint becomes weaker at larger $n$.
For certain reasonable choices of basis at two and three loops the two constraints in unison appear strong enough
to determine MHV integrands uniquely for all $n$.
}

\begin{document}
\preprint{Brown-HET-1625}
\maketitle
\pagebreak

\section{Introduction}

In the rapidly developing field of scattering amplitudes in maximally
supersymmetric Yang-Mills (SYM) theory it sometimes happens that a
new result obtained at enormous cost in time and effort can surprisingly
soon thereafter be rederived by a different
method, or at least understood in terms
of a new framework, as progress marches inexorably on.
This is in no way an indication of wasted effort, as
it is precisely the availability of expensive but valuable new
`theoretical data' which is crucial for inspiring the insights which
in turn allow the next generation of data to be collected.

This bootstrap paradigm suggests that it is never inappropriate to
reexamine previously obtained calculational results to look for
signs of underlying structure or properties which could provide hints
for helping to guide the journey forward.  Our focus in this paper
is on integrands for planar multi-loop MHV amplitudes in SYM theory.
In principle there is no conceptual obstacle to constructing these objects
using a variety of reasonably efficient techniques, including generalized
unitarity~\cite{Bern:1994zx,Bern:1994cg,Buchbinder:2005wp,Bern:2007ct,Cachazo:2008dx},
integrand-level recursion
relations~\cite{CaronHuot:2010zt,ArkaniHamed:2010kv,Boels:2010nw}, or
the equivalence between SYM integrands and
supersymmetric Wilson
loops~\cite{Mason:2010yk,CaronHuot:2010ek,Belitsky:2011zm,Adamo:2011pv}
or correlation
functions~\cite{Alday:2010zy,Eden:2010zz,Eden:2010ce,Eden:2011yp,Eden:2011ku,Eden:2011we,Eden:2012tu}.

Specifically our focus is to investigate the extent to which the results
of~\cite{Bourjaily:2011hi} generalize to $n>4$-particle amplitudes.
In that paper it was
shown through seven loops, and conjectured to be true to all loop order,
that the
4-particle planar integrand of the logarithm in SYM theory
has mild behavior (the leading divergence vanishes) in a collinear limit.
In this paper we reformulate the condition of~\cite{Bourjaily:2011hi}
in a very simple way (see equation~(\ref{eq:colliconst}) below) which no longer makes
any reference to the logarithm, and we verify that it holds for
two and three-loop MHV integrands for $n>4$.

Furthermore in~\cite{Bourjaily:2011hi} it was conjectured (and again, shown
through seven loops) that the 4-particle planar integrand
is uniquely determined by simultaneously
imposing dual conformal invariance, dihedral symmetry, and the mild
collinear behavior of the logarithm.
For $n>4$ it is almost immediately seen that this collinear limit by itself cannot
be enough to determine integrands (and is actually a relatively weak constraint
for large $n$), so we will add to our arsenal a second constraint:  that the
$n$-particle integrand must reduce smoothly to the $n-1$-particle integrand
in the soft limit, i.e. as the momentum of particle $n$ is taken to zero.
This condition, which was pointed out in~\cite{Drummond:2010mb}, was not discussed
in~\cite{Bourjaily:2011hi} as it is trivial for four-particle
integrands.  Here we find that the soft constraint is interestingly
complementary to the collinear constraint, becoming more powerful at larger $n$.

Before proceeding too far let us address an immediate comment one might have on our
approach.  Certainly there are various well-known but non-trivial integrat{\it ed}
quantities, such as for $n=6$
the MHV remainder function~\cite{Bern:2005iz} or the NMHV ratio
function~\cite{Drummond:2008vq}
(both of which have been explicitly evaluated,
see~\cite{DelDuca:2009au,Goncharov:2010jf}
and~\cite{Dixon:2011nj} respectively), which
vanish in any soft or collinear limit.  However let us emphasize that the `collinear'
limit we discuss here is one in which a loop integration variable, not an external
momentum, becomes collinear with another external momentum.  This type of collinear
limit can therefore only be probed at the level of an integrand, not on an
integrated quantity.  Moreover the MHV remainder function $R_n^{(L)}$
has no known integrand---that is,
there is no known rational function of external data and
$L$ loop integration variables which, when integrated, gives precisely
$R_n^{(L)}$.
It would be extremely interesting if such an objected could be
found, or even better an algorithm given showing how to determine it for any $n$ and
$L$ in terms of the relevant amplitude integrands (which are explicitly known
at two and three loops~\cite{ArkaniHamed:2010gh}).

In order to apply the soft and collinear constraints towards the
generation of integrands it is necessary to first choose a basis of objects in terms
of which one believes a certain integrand should be expressible.
The soft+collinear constraints then become a system of linear equations
on the coefficients in that basis.
The analysis carried out in~\cite{Bourjaily:2011hi} for $n=4$ benefitted greatly
from an especially manageable basis,
the collection of dual conformally invariant
four-point diagrams.
In particular it seems
(and was proven through seven loops) that this collection is linearly independent
at the level of the integrand,
which is certainly not true of the collection of
natural $n>4$-point dual conformally invariant diagrams which one can easily write
down.

At two loops, the construction of a basis $\mathcal{B}_2$
for general planar Feynman integrals
has been discussed in~\cite{Gluza:2010ws}. It would be fascinating
to study the strength of the collinear and soft limits in the subset
$\mathcal{D}_2
\subset \mathcal{B}_2$ of dual conformally invariant integrals, but we feel
this very general analysis is still beyond our reach at the moment.
Instead
we will take a `Goldilocks' approach to the basis:
first we try small bases at two and three loops which are just
barely large enough
to be known~\cite{ArkaniHamed:2010gh} to be able to express the MHV
amplitude for all $n$.  Encouragingly we find that in these cases it is
true that the coefficient of every term in the integrand is uniquely determined
by the soft+collinear constraints.
Then at two loops and for relatively small $n$ we investigate the strength of
the soft+collinear constraints in two other bases: a `too large' basis composed
of arbitrary dual conformally invariant
rational functions of momentum-twistor four-brackets, where
not surprisingly the constraints are too weak,
and finally a `just right' subset
of the latter in which we find that the soft+collinear constraints appears
precisely
powerful to uniquely determine integrands for all $n$.

\section{Setup}

\subsection{Conventions}
We begin by making a potentially jarring notational change: for the rest of this paper, we will refer to ``the number of loops" as $\l$ instead of $L$. We make this change in order to denote the planar $n$-particle $\l$-loop MHV integrand\footnote{Our convention for the overall
normalization of integrands agrees with that of~\cite{ArkaniHamed:2010gh}
and differs slightly from that of~\cite{Bourjaily:2011hi}:  the latter
$\l$-loop integrands are a factor of $\l!$ larger than the former.} by $M^{(\l)}_n$, and the corresponding integrand of the logarithm by $L^{(\l)}_n$.
More details about these quantities, and the exact relation between them,
are reviewed in~\cite{Bourjaily:2011hi}.
They are rational functions of the momentum-twistor
four-brackets~\cite{Hodges:2009hk}
involving external and loop integration dual momentum variables.
The former are denoted by $Z_i$ for $i=1,\ldots,n$ while the latter, being
off-shell, need to be described
by a pair of momentum twistors which we denote by $Z_{A_i}, Z_{B_i}$ for
$i=1,\ldots,\l$.
Occasionally we will use the notation $M^{(\l)}_n(1,2,\ldots,n)$ or
$M^{(\l)}_n[1,\ldots,\ell]$ when it is necessary to make explicit reference
to the dependence of an integrand on the external or internal variables
respectively.  In examples through three loops we will typically use
the loop integration variables $Z_A,Z_B,Z_C,Z_D,Z_E$ and $Z_F$.

\subsection{Constraints}

Constraints define how the integrand $M^{(\l)}_n$ behaves when the external and/or internal variables are taken to certain limits. In this paper we will be probing the behavior of integrands under two different limits: soft and collinear.

The soft limit involves taking the limit where one of the external momenta vanishes: $p_n\to 0$. In the language of momentum twistors, the soft limit takes the form
\begin{equation}\label{eq:softlimit}
Z_n\to \alpha\, Z_1 + \beta\, Z_{n-1}
\end{equation}
for arbitrary $\alpha,\, \beta$. It was first noted in~\cite{Drummond:2010mb} that the integrand behaves very nicely under the soft limit, in particular the $n$-particle integrand reduces directly to the $(n-1)$-particle integrand when $p_n \to 0$. For the purpose of this paper we will formalize this as the {\it soft constraint}:\\ \\
\noindent{\bf \underline{Soft Constraint}}\\ \noindent{\it The integrand of the $n$-particle $\l$-loop amplitude
in planar SYM theory behaves as}
\begin{equation}
M^{(\l)}_n(1,2,\ldots,n-1,n)\,\to\,M^{(\l)}_{n-1}(1,2,\ldots,n-1)
\end{equation}
{\it in the limit~(\ref{eq:softlimit}) for all $n$ and $\l$.}\\ \\
Of course it is to be understood that $M^{(\l)}_n = 0$ for $n<4$, $\l>0$.

The collinear constraint is based on the observation that the {\it logarithm} of an amplitude has softer IR-divergences than expected when an internal momenta becomes collinear with an external momenta. This softness is well-understood at the level of the integral, but has only been recently discovered to be a property of integrands~\cite{Bourjaily:2011hi}. We emphasize that this was an empirical discovery, and there is no rigorous proof that the property must hold in general.  The collinear limit we will be considering in this paper is (see~\cite{Bourjaily:2011hi} for more discussion)
\begin{equation}\label{eq:collinearlimit1}
Z_{A_1} \to Z_2+\O(\epsilon), \qquad Z_{B_1} \to  Z_1 +   Z_3\,+\O(\epsilon)\,.
\end{equation}
Naturally this limit will produce infrared divergences, and at the level of the integrand these manifest as $1/\e^2$ and $1/\e$ poles in the limit $\e\to0$. However, we conjecture that the $1/\e^2$ pole in $L^{(\l)}_n$ vanishes. We phrase this explicitly as the {\it collinear constraint}: \\ \\
\noindent{\bf \underline{Collinear Constraint}}\\
\noindent
{\it The integrand of the logarithm of the $n$-particle $\l$-loop amplitude
in planar SYM theory behaves as}
\begin{equation}
\label{eq:logcoll}
L^{(\l)}_n[1,2,\ldots,n-1,n]\,\to\,\frac{0}{\epsilon^2} + \O(1/\epsilon)
\end{equation}
{\it in the limit~(\ref{eq:collinearlimit1}) for all $\l>1$}.\\ \\
We have verified this conjecture by explicit
calculation through seventeen points at two loops and eleven
points at three loops.

\subsection{Collinear Redux}
The collinear constraint in the previous section describes how the integrand of the logarithm behaves under the collinear limit. However, there is a more direct way of framing the collinear constraint in terms of the integrand of the amplitude itself. First, let us briefly reintroduce the collinear limit in a more explicit way:
\begin{equation}\label{eq:collinearlimit}
Z_{A_1} \to Z_2+\e Z_X, \qquad Z_{B_1} \to  \alpha Z_1 + \beta Z_2 +\gamma Z_3,
\end{equation}
where $Z_X$ is an arbitrary four-vector describing the path taken as $\e\to0$ and the $\alpha,\beta,\gamma$ terms, with $\alpha$ and $\beta$ non-zero, are included in the interest of complete generality. To understand how integrands behave under the collinear limit, we will first consider the $n$-particle one-loop integrand
\begin{equation}\label{eq:oneloop}
M^{(1)}_n =\sum_{i<j} \frac{\ab{AB\,(i\smm1\,i\,i\smp1)\newcap(j\smm1\,j\,j\smp1)}\ab{ijn1}}{\ab{AB\,i\smm1\,i}\ab{AB\,i\,i\smp1}\ab{AB\,j\smm1\,j}\ab{AB\,j\,j\smp1}\ab{ABn1}}.
\end{equation}
Under the collinear limit~(\ref{eq:collinearlimit}), the only $\O(1/\e^2)$ pole comes from the $i=2, j=3$ term. This can be easily evaluated to give
\begin{equation}\label{eq:oneloopresult}
\left(M^{(1)}_n\right)_{\rm collinear}=-\frac{1}{(\e\alpha\gamma\ab{123X})^2} + \O(1/\e).
\end{equation}
With this explicit example worked out, we will drop future factors of $\alpha, \gamma$, and $\ab{123X}$ for notational clarity. They can be restored in any formula by replacing $\e^2$ with $(\e\alpha\gamma\ab{123X})^2$.

Direct manipulation of the general $n$-particle two-loop integrand is more difficult, however, we can simplify matters considerably by employing the collinear constraint on the integrand of the logarithm~(\ref{eq:colliconst1}). The two-loop integrand of the logarithm is given in terms of the one- and two-loop integrands by
\begin{equation}
L^{(2)}_n[1,2] = M^{(2)}_n[1,2]-\frac{1}{2}M^{(1)}_n[1]M^{(1)}_n[2].
\end{equation}
Taking the collinear limit and employing~(\ref{eq:oneloopresult}) gives
\begin{equation}
L^{(2)}_n[1,2] \to \left(M^{(2)}_n[1,2]\right)_{\rm collinear}+\frac{1}{2\e^2}M^{(1)}_n[2]+\O(1/\e).
\end{equation}
Imposing the collinear constraint then requires that
\begin{equation}
\left(M^{(2)}_n[1,2]\right)_{\rm collinear}=-\frac{1}{2\e^2}M^{(1)}_n[2] +\O(1/\e).
\end{equation}

A similar calculation, worked out in detail in appendix A, shows that imposing the collinear constraint at three loops is equivalent to requiring
\begin{equation}
\left(M^{(2)}_n[1,2,3]\right)_{\rm collinear}=-\frac{1}{3\e^2}M^{(2)}_n[2,3] +\O(1/\e).
\end{equation}
By now a likely pattern has emerged,
and indeed we prove in appendix A that the logarithm-level collinear
constraint~(\ref{eq:logcoll}) is exactly equivalent, to all loop order,
to the following amplitude-level {\it collinear constraint}:\\ \\
\noindent{\bf \underline{Collinear Constraint} (redux)}\\
\noindent{\it The integrand of the $n$-particle $\l$-loop amplitude
in planar SYM theory behaves as
\begin{equation}\label{eq:colliconst}
M^{(\l)}_n[1,2,\ldots,\l]\,\to\,-\frac{1}{\l} M^{(\l-1)}_{n}[2,3,\ldots,\l]
\end{equation} at $\O(1/\e^2)$ in the limit~(\ref{eq:collinearlimit1}) for all $\l>0$.}\\ \\
We make no attempt here to prove that this is indeed a necessary property of SYM theory amplitudes (a very nice argument in this direction has recently been discovered by~\cite{CHH}). Instead, we conjecture that it is true in general and have verified it to be true at three loops through eleven particles. For the rest of the paper the phrase ``collinear constraint" will refer to~(\ref{eq:colliconst}). The benefit of defining the collinear constraint in this way is not that it is stronger or more informative than the previous definition.  Rather, it is computationally and conceptually more simple while containing the same information. Furthermore, this new definition highlights the fact that the collinear and soft limits induce complementary behavior on the same object $M_n^{(\l)}$.

\section{Procedure}
Now that we have defined our constraints, we will describe in detail how they can be used to construct integrands. We begin by postulating a suitable basis of integrands, which will ideally be general enough to describe our desired amplitude while also being small enough to be computationally feasible. We then try to find a linear combination of these basis integrands that satisfy both the collinear and soft constraints.
\subsection{Defining a Basis}\label{sec:basis}
Constructing a basis for the $n$-particle $\l$-loop integrand is a multi-step process, briefly outlined here:
\begin{enumerate}
\item Select a {\it seed} for the basis.
\item Use the seed to generate a {\it pre}-basis.
\item Impose symmetries and linear independency on the pre-basis to determine the basis.
\end{enumerate}
The underlying theme here is to consider as large of a basis as possible, with the hope that the power of our constraints will cast aside all but the small sub-set of desired basis terms.
For example, in a later section, we will consider a basis of general rational functions of momentum-twistor four-brackets, abandoning the notion of a ``diagram" completely.

The choice of seed determines what type of rational functions will enter into our final basis. For example, at two-loops we will first consider the seed
\vspace{-0.2cm}
\begin{equation}
\nonumber\hspace{-0.6cm}\figBox{0}{-1.82}{0.55}{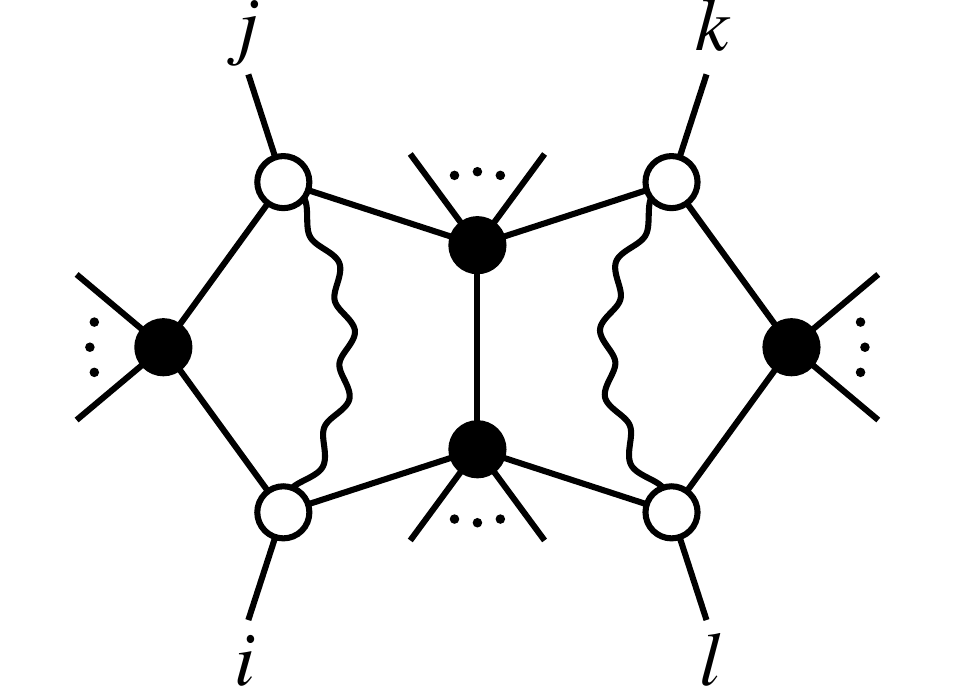}\!\!\!\!\!=\left\{\begin{array}{l}
\displaystyle\frac{\ab{AB\,(i\smm1\,i\,i\smp1)\newcap(j\smm1\,j\,j\smp1)}\ab{i\,j\,k\,l}}{\ab{AB\,i\smm1}\ab{AB\,i\,i\smp1}\ab{AB\,j\smm1\,j}\ab{AB\,j\,j\smp1}\ab{AB\,CD}}\\\displaystyle\times\frac{\ab{CD\,(k\smm1\,k\,k\smp1)\newcap(l\smm1\,l\,l\smp1)}}{\ab{CD\,k\smm1k}\ab{CD\,k\,k\smp1}\ab{CD\,l\smm1\,l}\ab{CD\,l\,l\smp1}}\end{array}\right\}\label{MHVInt}
\end{equation}
This particular diagram will be discussed in more depth in the two-loop section, for now we are just using it as an example. We define the pre-basis as the collection of {\it all} iterations of the seed. When dealing with diagrams we also of course restrict ourselves to only planar diagrams. For more general, non-diagrammatic bases considered in later sections we will have no notion of ``planar" and ``non-planar." For this particular seed, we will consider $\{i,j,k,l\}$ ranging over all possible planar orderings for a given $n$, so there will be a total of $4 {n \choose 4}$ elements in the pre-basis.

It is well-known that the amplitude must obey a dihedral symmetry $D_n$, composed of cyclic and mirror symmetries, in the external variables:
\begin{equation}
\mathcal{A}(1,2,\ldots,n) = \mathcal{A}(n,1,2,\ldots,n-1)=\mathcal{A}(n,n-1,\ldots,2,1)
\end{equation}
as well as an $\mathfrak{s}_{\l}$ permutation symmetry in the $\l$-integration variables. We can forcibly impose these symmetries by summing each member of the pre-basis over all of its $D_n \times \mathfrak{s}_{\l}$ images, keeping of course each uniquely generated sum only once. This will give
\begin{equation}\label{eq:prebasis}
\text{pre-basis}=\left\{{\cal I}_1,{\cal I}_2,\ldots,{\cal I}_m\right\}
\end{equation}
where the ${\cal I}_i$ are each $D_n \times \mathfrak{s}_\l$-invariant sums over pre-basis terms. At this stage we must also take into account any linear dependencies, such as Schouten identities, that may exist between elements of the pre-basis. We therefore want to eliminate elements in our pre-basis that can be decomposed into a linear combination of other elements. Computationally, this can be achieved by evaluating~(\ref{eq:prebasis}) over $m$ sets of random values $\{r_i\}$ of the kinematic and loop variables to generate an $m\times m$ matrix:
\begin{equation}
 \begin{pmatrix}
{\cal I}_1(r_1) &{\cal I}_2(r_1)  & \cdots &{\cal I}_m(r_1)  \\
{\cal I}_1(r_2) &{\cal I}_2(r_2)  & \cdots &{\cal I}_m(r_2)  \\
  \vdots  & \vdots  & \ddots & \vdots  \\
{\cal I}_1(r_m) &{\cal I}_2(r_m)  & \cdots &{\cal I}_m(r_m)  \\
 \end{pmatrix}
\end{equation}
This can then be row-reduced to expose a set of $k\le m$ linearly independent basis elements.
(Of course there is always the danger that some of the randomly selected values may
accidentally indicate some apparent linear dependencies which are not truly present.
To guard against this one can in practice choose far more than only $m$ sets
of different random values.)

We then define our basis as the collection of linearly independent,  $D_n \times \mathfrak{s}_{\l}$-symmetric rational functions resulting from this procedure:
\begin{equation}\label{eq:prebasis2}
\text{basis}=\left\{{\cal I}_1,{\cal I}_2,\ldots,{\cal I}_k\right\}.
\end{equation}
{}From such a basis one can then construct an ansatz
\begin{equation}\label{eq:ansatz}
M_n^{(\l)}=\sum_i c_i \,{\cal I}_i.
\end{equation}
Our goal is now to determine the collection $\{c_i\}$ of numerical constants which will give us the correct integrand.
\subsection{Imposing Constraints}
Once we have a fully symmetrized and linearly independent basis, the procedure for obtaining the $\l$-loop integrand $M_n^{(\l)}$ from the collinear constraint can be carried out as follows:

\begin{enumerate}
\item Compute the collinear limit of the $n$-point $\l$-loop ansatz~(\ref{eq:ansatz}), and collect from each ${\cal I}$ the $\O(1/\epsilon^2)$ contribution; call these contributions $c_i({\cal I}_i)_{\mathrm{collinear}}$;
\item We now wish to determine the collection of numerical constants $\{c_i\}$ such that \[ \sum_i c_i({\cal I}_i)_{\mathrm{collinear}}=-\frac{1}{\l}M^{(\l-1)}_n.\]  By repeatedly evaluating this equation at sufficiently many random independent values of the remaining variables (i.e., $(Z_{A_2},Z_{B_2}), \ldots, (Z_{A_\ell},Z_{B_\ell})$
as well as $Z_1,\ldots,Z_n)$, this equation can be turned into a linear system on the $c_i$ with constant coefficients.
\end{enumerate}
The soft constraint can be imposed in an exactly analogous way, except we are now determining $c_i$ that satisfy
\[\sum_i c_i({\cal I}_i)_{\mathrm{soft}}=M_{n-1}^{(\l)}.\]
and the remaining variables are $(Z_{A_1},Z_{B_1}),\ldots,(Z_{A_\ell},Z_{B_\ell})$ as well as
$Z_1,\ldots,Z_{n-1})$.

It should be noted that, depending on the choice of basis, $({\cal I}_i)_{\mathrm{collinear}}$ and $({\cal I}_i)_{\mathrm{soft}}$ may vanish for certain choices of $i$. If, for example,  $({\cal I}_1)_{\mathrm{collinear}}=0$ , the collinear limit will be blind to ${\cal I}_1$ and will provide no constraint on the coefficient $c_1$. Throughout the rest of this paper we will be essentially be asking the same questions: for a given $n$-particle, $\l$-loop ansatz,
\begin{itemize}
\item how many of the $c_i$ are fixed by the collinear constraint alone?
\item how many of the $c_i$ are fixed by the soft constraint alone?
\item how many of the $c_i$ are fixed by the collinear and soft constraints in conjunction?
\end{itemize}
Answering these questions over a range of bases will give us some insight into how strong these constraints are at various $n$ and $\l$.

\section{Two-Loops}\label{sec:twoloop}
An important first step in employing the collinear and soft constraints is to verify that they reproduce previously known integrands. In~\cite{ArkaniHamed:2010gh}, the $n$-particle MHV integrand was given at two- and three-loops. These representations are built upon a special set of integrands which are chiral and have unit leading singularities. Our first goal, then, is to use these integrands to generate a basis using these integrands and see if the collinear and soft constraints can reproduce the known integrand. At two-loops, the $n$-particle integrand is given by

\vspace{-0.4cm}
\begin{equation}\label{eq:twoloopintegrand}
M_{n}^{(2)}=\displaystyle\underset{\substack{i<j<k<l<i}}{\frac{1}{2}\text{{\Huge$\sum$}}\phantom{\frac{1}{2}\!\!}}\hspace{-0.2cm}\raisebox{-1.6cm}{\includegraphics[scale=0.5]{two_loop_integrand.pdf}}
\end{equation}
with
\vspace{-0.1cm}
\begin{equation}\label{eq:twoloopbasis}
\hspace{-1cm}\figBox{0}{-1.82}{0.5}{two_loop_integrand.pdf}\!\!\!\!\!=\left\{\begin{array}{l}
\displaystyle\frac{\ab{AB\,(i\smm1\,i\,i\smp1)\newcap(j\smm1\,j\,j\smp1)}\ab{i\,j\,k\,l}}{\ab{AB\,i\smm1}\ab{AB\,i\,i\smp1}\ab{AB\,j\smm1\,j}\ab{AB\,j\,j\smp1}\ab{AB\,CD}}\\\displaystyle\times\frac{\ab{CD\,(k\smm1\,k\,k\smp1)\newcap(l\smm1\,l\,l\smp1)}}{\ab{CD\,k\smm1k}\ab{CD\,k\,k\smp1}\ab{CD\,l\smm1\,l}\ab{CD\,l\,l\smp1}}\end{array}\right\}
\end{equation}
We will start by taking~(\ref{eq:twoloopbasis}) as the seed of our basis, and then impose constraints in the hope of reproducing~(\ref{eq:twoloopintegrand}).

Boundary terms occur when $j=i+1$ and/or $l=k+1$, in which case one or both of the wavy-line numerators cancels out a propagator, resulting in a box rather than pentagon topology. As we will see, these boundary terms will be of fundamental importance with regards to the collinear limit, so we write them down explicitly here:
\begin{equation}
\hspace{-1.85cm}\figBox{0}{-1.7}{0.45}{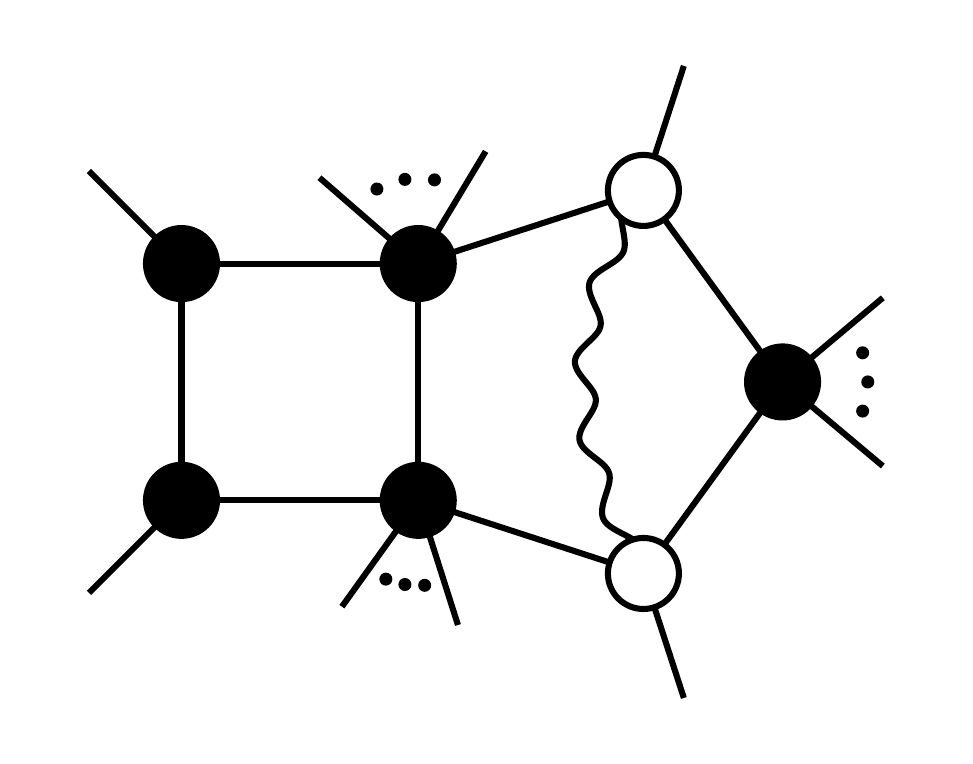}\;\figBox{0}{-1.7}{0.45}{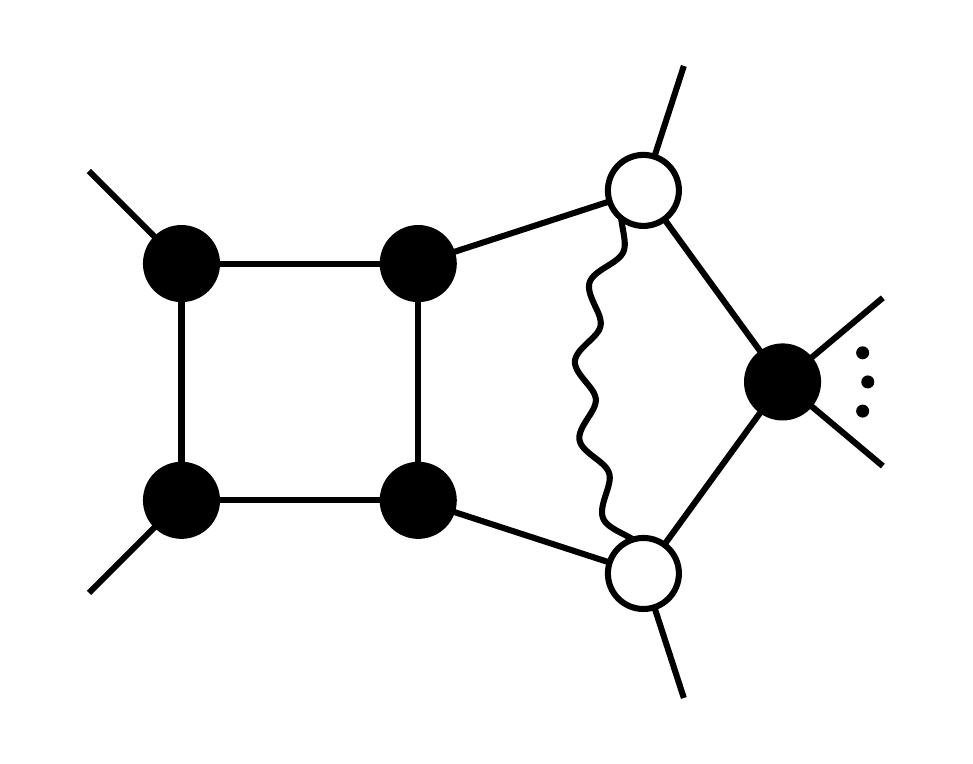}\;\figBox{0}{-1.15}{0.45}{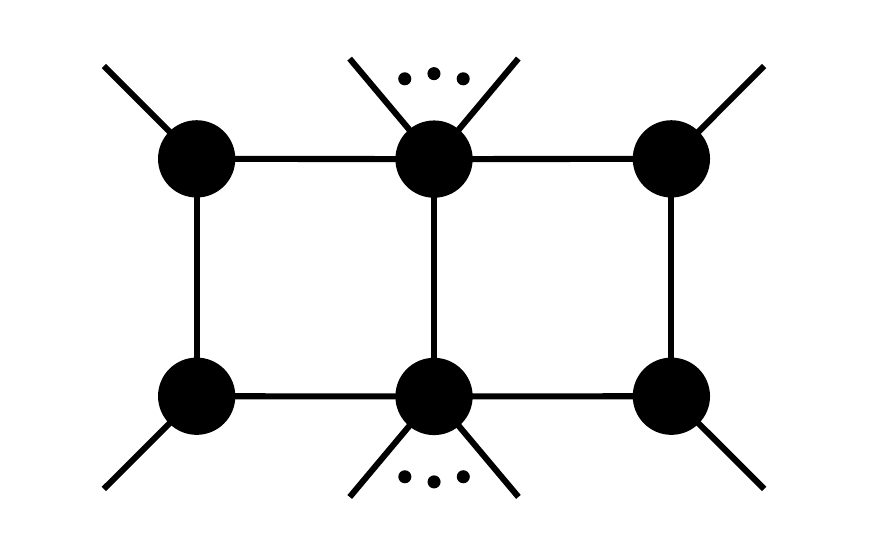}\hspace{-2cm}\label{MHVInt2}
\end{equation}

\subsection{Five- and Six-Point Examples}

With the two-loop basis, imposing the dihedral and loop-variable symmetries is equivalent to selecting topologically distinct diagrams. At $n=5$, there are only two non-isomorphic diagrams:

\begin{equation}
\hspace{-1.85cm}\figBox{0}{-1.7}{0.45}{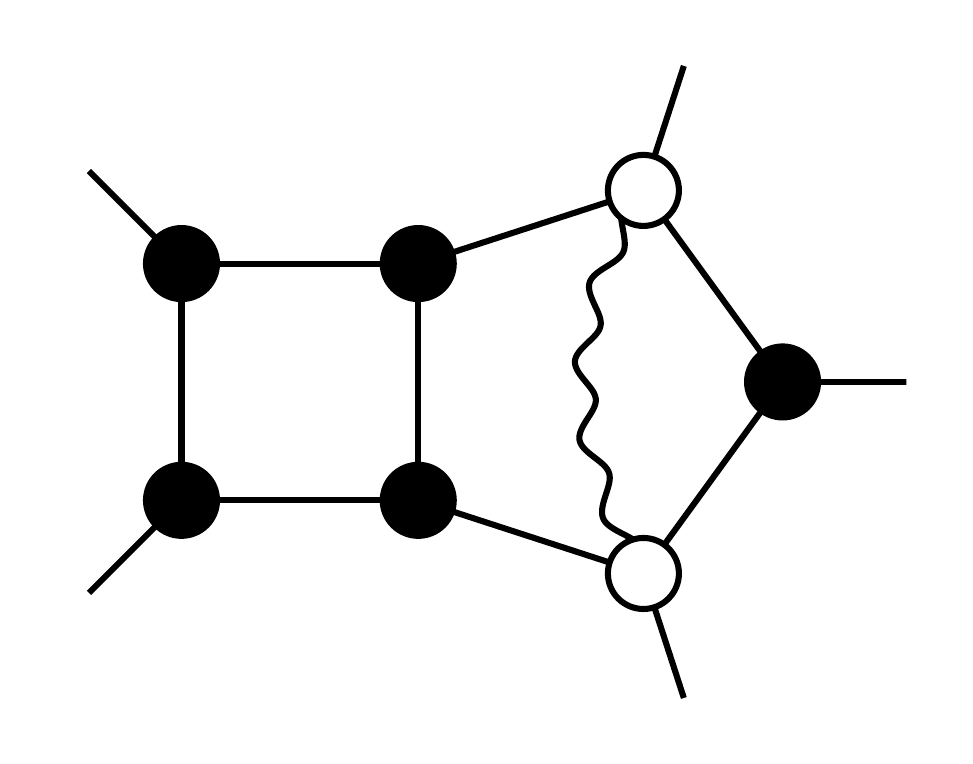}\;\figBox{0}{-1.15}{0.45}{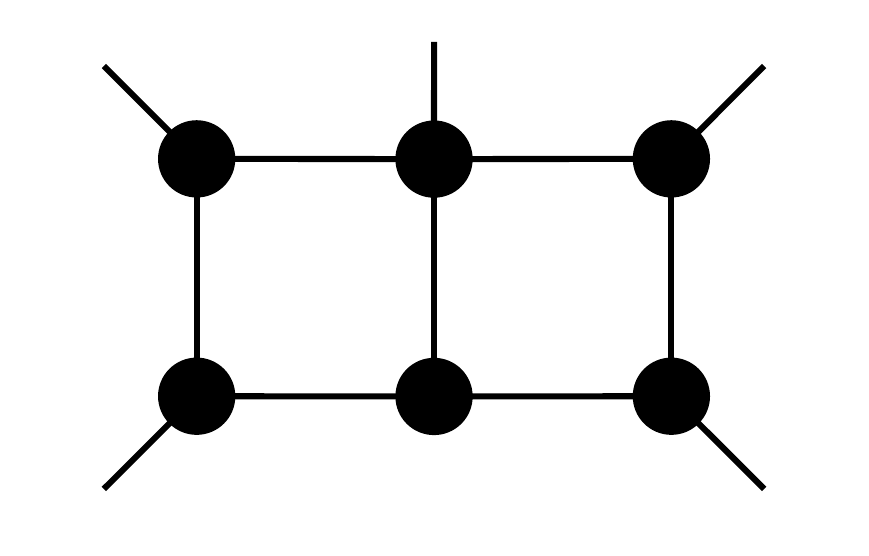}\hspace{-2cm}
\end{equation}
We therefore have the ansatz
\begin{equation}
M_5^{(2)}= a\,{\cal I}_1^{(2)}+b\,{\cal I}_2^{(2)}
\end{equation}
where each ${\cal I}^{(2)}_i$ is a sum of ten diagrams related by symmetry:
\begin{equation}
\begin{split}
&\hspace{-.2cm}{\cal I}_1^{(2)}=\frac{\ab{1235}^2 (\ab{1245}\ab{CD34}-\ab{1345} \ab{CD24})}{\ab{AB12}\ab{AB23}\ab{AB51} \ab{ABCD} \ab{CD23}\ab{CD34} \ab{CD45}\ab{CD51}}\,+\,\text{sym.} \\ \\
&\hspace{-.2cm}{\cal I}_2^{(2)}=\frac{\ab{1235}\ab{1245}\ab{1345}}{\ab{AB12}\ab{AB23}\ab{AB51}\ab{ABCD}\ab{CD34}\ab{CD45}\ab{CD51}}\,+\,\text{sym.}
\end{split}
\end{equation}

Both ${\cal I}_1$ and ${\cal I}_2$ clearly have non-zero $\O(1/\e^2)$ contributions under the collinear limit. We can therefore assemble the equation
\begin{equation}
 a\left({\cal I}_1^{(2)}\right)_{\rm collinear}+b\left({\cal I}_2^{(2)}\right)_{\rm collinear} = -\frac{1}{2}M^{(1)}_5.
 \end{equation}
After generating a system of equations on $a$ and $b$ by evaluating this equation at random kinematical values, we see that the only solution is $a = b = 1/2$, in accordance with the known value for the two-loop five-particle integrand.

A similar story occurs with the soft limit: both diagrams give non-zero, linearly independent contributions under the soft limit, and so the soft limit constraint fixes $a=b=1/2$.

At $n=6$ there are five topologically distinct diagrams:
\begin{equation}
\begin{split}
\hspace{-1.85cm}\figBox{0}{-1.7}{0.5}{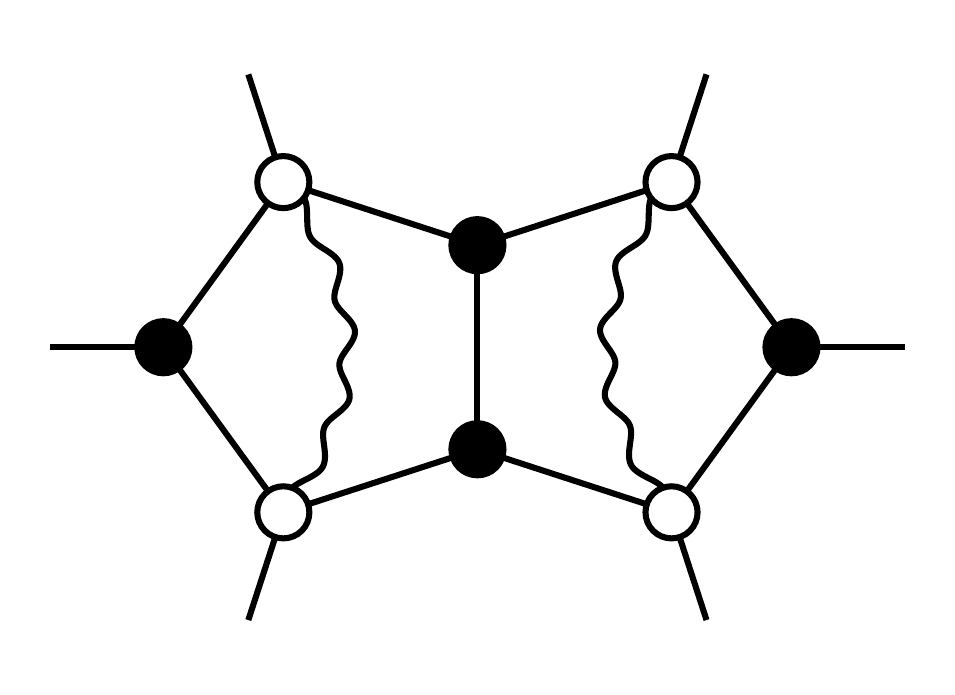}\;\figBox{0}{-1.7}{0.45}{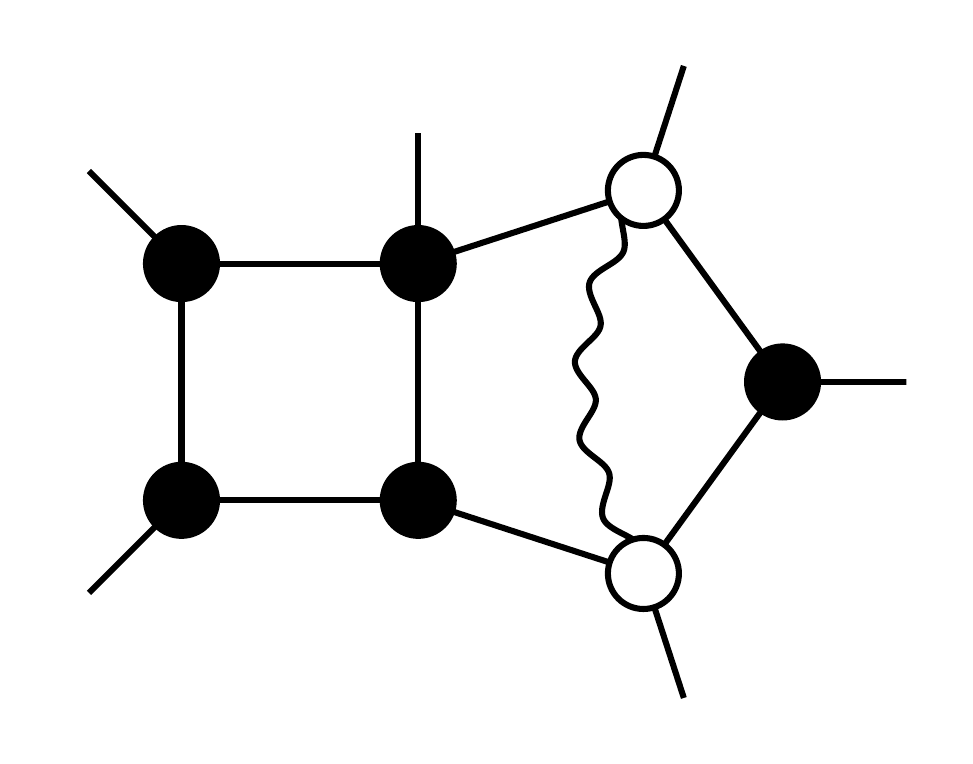}\;\figBox{0}{-1.7}{0.45}{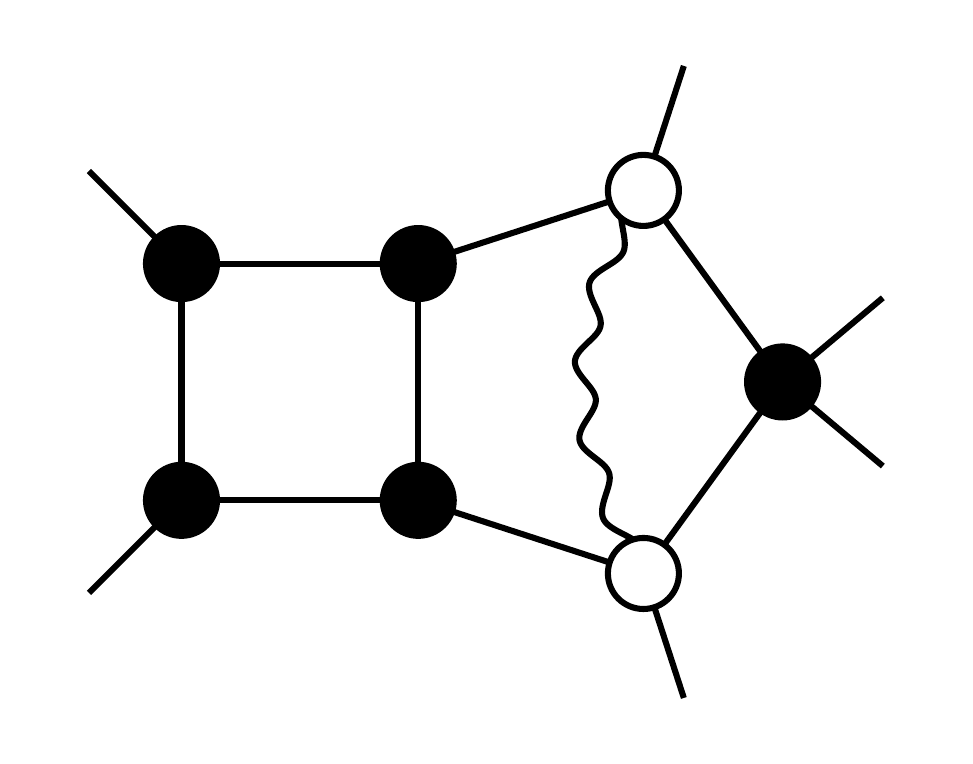}\hspace{-2cm}\\
\hspace{-3cm}\figBox{0}{-1.15}{0.45}{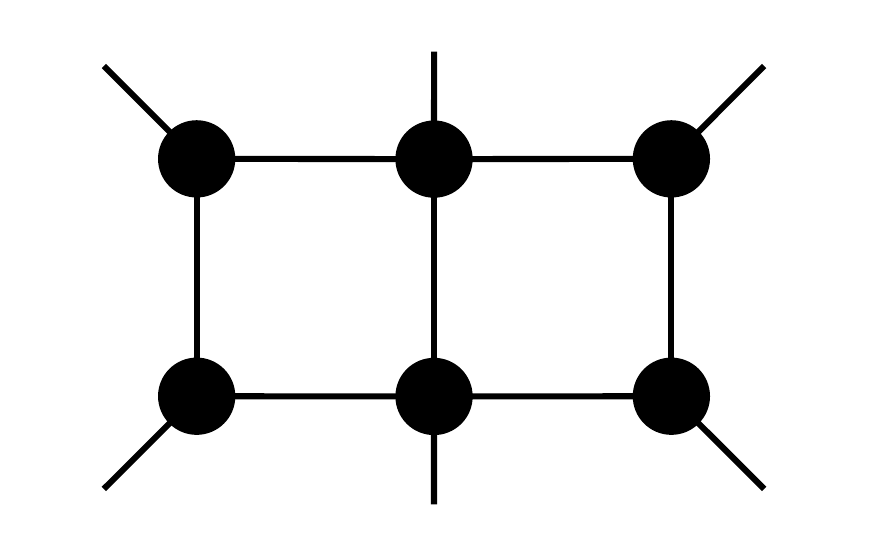}\;\figBox{0}{-1.15}{0.45}{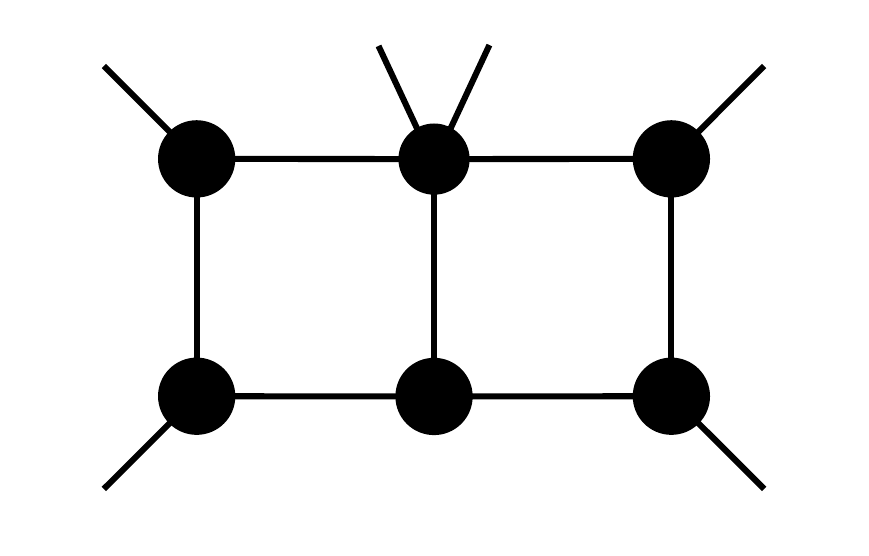}
\end{split}
\end{equation}

Under the collinear limit, the double-pentagon topology does {\it not} have a contribution at $\O(1/\e^2)$. The other topologies, all featuring at least one box, have $\O(1/\e^2)$ divergences under the collinear limit. Furthermore, the $\O(1/\e^2)$-poles from each diagram are linearly independent rational functions. Therefore, the collinear limit constraint determines four of the five coefficients in the ansatz.

All of the topologies survive the soft limit and are linearly independent, so the integrand is completely determined by the soft limit constraint.
\subsection{General Results}
The size of the basis and the number of coefficients fixed by the collinear constraint are summarized through $n=17$ in table~\ref{table:twoloop}.

The soft constraint is not included in table~\ref{table:twoloop} because it fixes all of the coefficients through $n=17$, and we conjecture that it fixes all of the coefficients for general $n$. This is because the basis generated by~(\ref{eq:twoloopbasis}) is very well-behaved under the soft limit. Specifically, the $n$-particle integrand directly becomes the $(n-1)$-particle integrand term-by-term under the soft limit:
 \begin{displaymath}
   \{i,j,k,l\}_n \xrightarrow{p_n~ \rightarrow~ 0~} \left\{
     \begin{array}{lr}
       \{i,j,k,l\}_{n-1} & : n \notin \{i,j,k,l\}\\
       0 & : n \in \{i,j,k,l\}
     \end{array}
   \right.
\end{displaymath}
(where the notation $\{i,j,k,l\}_n$ describes a particular diagram in the $n$-particle basis). Therefore, each basis term will give a non-zero contribution under the soft limit. As we have explicitly verified through $n=17$, these contributions remain linearly independent and so all are fixed by the soft constraint.
\begin{table}[h]
\begin{center}
  \begin{tabular}{ c|c|c|}
        \cline{2-3}
    & \multicolumn{2}{|c|}{{\bf \# of unfixed coefficients}}\\
     \hline
    \multicolumn{1}{|c|}{${\mathbf n}$} & Symmetrized basis & After collinear constraint \\ \hline
    \multicolumn{1}{|c|}{5}& 2 & 0   \\ \hline
    \multicolumn{1}{|c|}{6}  & 5 & 1 \\ \hline
    \multicolumn{1}{|c|}{7} & 8 & 2    \\ \hline
    \multicolumn{1}{|c|}{8} & 14 & 5    \\ \hline
    \multicolumn{1}{|c|}{9}  & 20 & 8   \\ \hline
    \multicolumn{1}{|c|}{10} &  30 & 14   \\ \hline
    \multicolumn{1}{|c|}{11} & 40 & 20   \\ \hline
    \multicolumn{1}{|c|}{12}  & 55 & 30 \\ \hline
    \multicolumn{1}{|c|}{13}  & 70  & 40   \\ \hline
    \multicolumn{1}{|c|}{14}  & 91 & 55   \\ \hline
    \multicolumn{1}{|c|}{15} & 112 &  70  \\ \hline
    \multicolumn{1}{|c|}{16} & 140 &  91   \\ \hline
    \multicolumn{1}{|c|}{17} & 168 &  112   \\ \hline
    \end{tabular}\caption{Results of the collinear constraint on the two-loop basis~(\ref{eq:twoloopbasis})}\label{table:twoloop}
    \end{center}
\end{table}

The collinear limit only fixes the coefficients of diagrams involving $\O(1/\e^2)$ divergences under the collinear limit. An interesting feature of the wavy-line numerator term of this basis is that it dampens IR divergences. In the collinear limit, diagrams that feature wavy-line numerators across both loops, i..e two pentagons (rather than two boxes or a box and a pentagon), will only have $\O(1/\e)$ divergences under the collinear limit and will not be fixed by the collinear constraint. Put more succinctly, the diagrams left unfixed by the collinear constraint are those without boxes. The question ``how many coefficients are left unfixed after imposing the collinear constraint" is equivalent to ``how many diagrams (up to isomorphism) involve two pentagons". In other words, imposing the collinear constraint is, topologically, equivalent to symmetrization with the added stipulation that there has to be at least one leg present at $x$ and $y$:
\vspace{-0.4cm}
\begin{equation}
\figBox{0}{-1.7}{0.5}{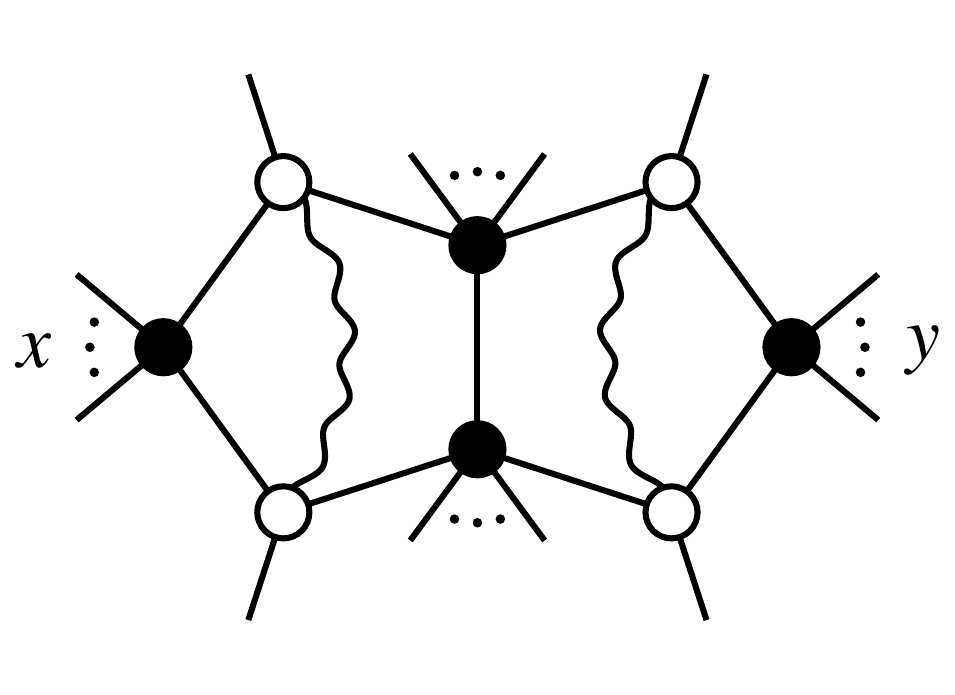}\vspace{-0.4cm}
\end{equation}
This leads to the relationship
\begin{equation}
\begin{split}
\text{\{\# of coefficients left unfixed after imposing the collinear constraint\}}(n)&\\
=\text{\{\# of topologies with only pentagons\}}(n)&\\
=\text{\{\# of topologies with pentagons and boxes\}}(n-2)&.
\end{split}
\end{equation}

The pattern that emerges in table~\ref{table:twoloop}, $\{0,1,2,5,8,14,\ldots\}$, has a number of possible interpretations, though none seem particularly insightful. The pattern can be interpreted as~\cite{OEIS}
\begin{itemize}
\item the number of aperiodic necklaces (Lyndon words) with 4 black beads and $n-4$ white beads.
\item the maximum number of squares that can be formed from $n$ lines.
\item partial sums of $\{1, 1, 3, 3, 6, 6, 10, 10,\ldots\}$.
\end{itemize}
It is interesting to note that, assuming this pattern holds for all $n$, the collinear constraint will asymptotically become useless at two-loops for large $n$. For odd $n$ the number of unconstrained coefficients left after imposing the collinear constraint goes as
\begin{equation}
\text{\{\# of unfixed coeff. after the collinear const.\}}(n_{{\rm odd}}) = \frac{(n-5)(n-3)(n-1)}{24}
\end{equation}
Therefore,
\begin{equation}
\frac{\text{\{\# of unfixed coefficients after the collinear const.\}}}{\text{\{\# of unfixed coefficients in the basis\}}}(n_{{\rm odd}})=\frac{n-5}{n+1}
\end{equation}
We can thus conclude that the percentage of coefficients fixed by the collinear limit in this particular basis at two-loops asymptotically approaches zero as $n$ goes to infinity.

\section{Three-Loops}\label{sec:threeloop}
The three-loop integrand given in~\cite{ArkaniHamed:2010gh} involves the pentagon and hexagon integrands with wavy-line numerators. The general form is given by
\vspace{-0.2cm}
\begin{equation}\label{eq:threeloopintegrand}
\hspace{-2cm}M_n^{(3)}=\!\!\!\!\displaystyle\underset{\substack{i_1\leq i_2<j_1\leq\\\leq j_2<k_1\leq k_2<i_1}}{\frac{1}{3}\!\text{{\Huge$\sum$}}\phantom{\frac{1}{2}\!\!}}\!\!\!\figBox{0}{-2.15}{0.55}{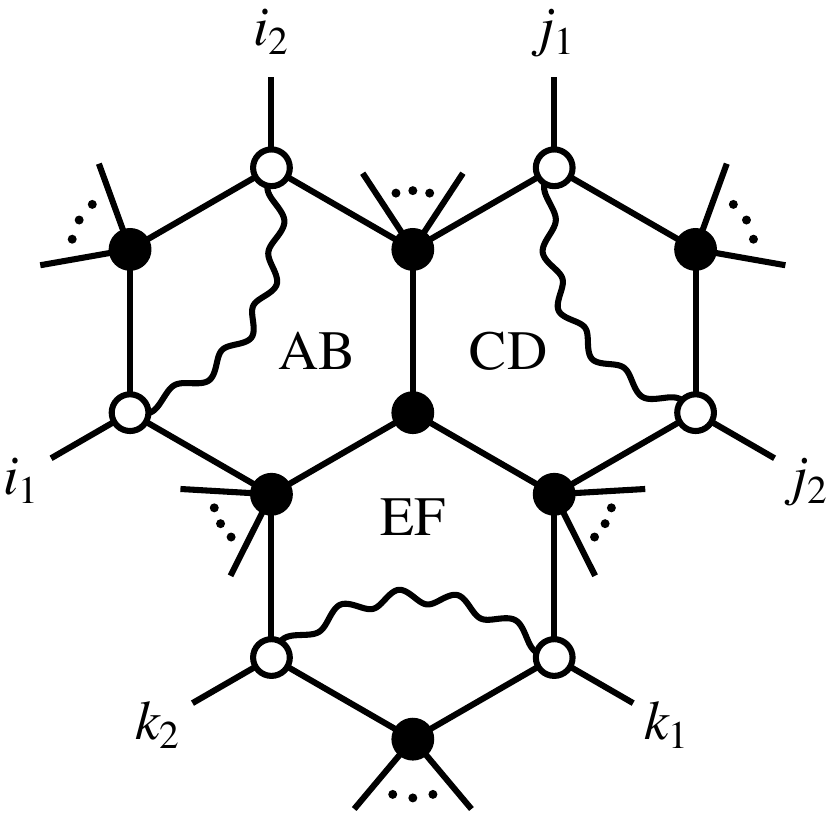}+\!\!\displaystyle\underset{\substack{i_1\leq j_1< k_1<\\< k_2\leq j_2< i_2<i_1}}{\frac{1}{2}\!\text{{\Huge$\sum$}}\phantom{\frac{1}{4}\!\!}}\!\!\!\!\figBox{0}{-1.8}{0.55}{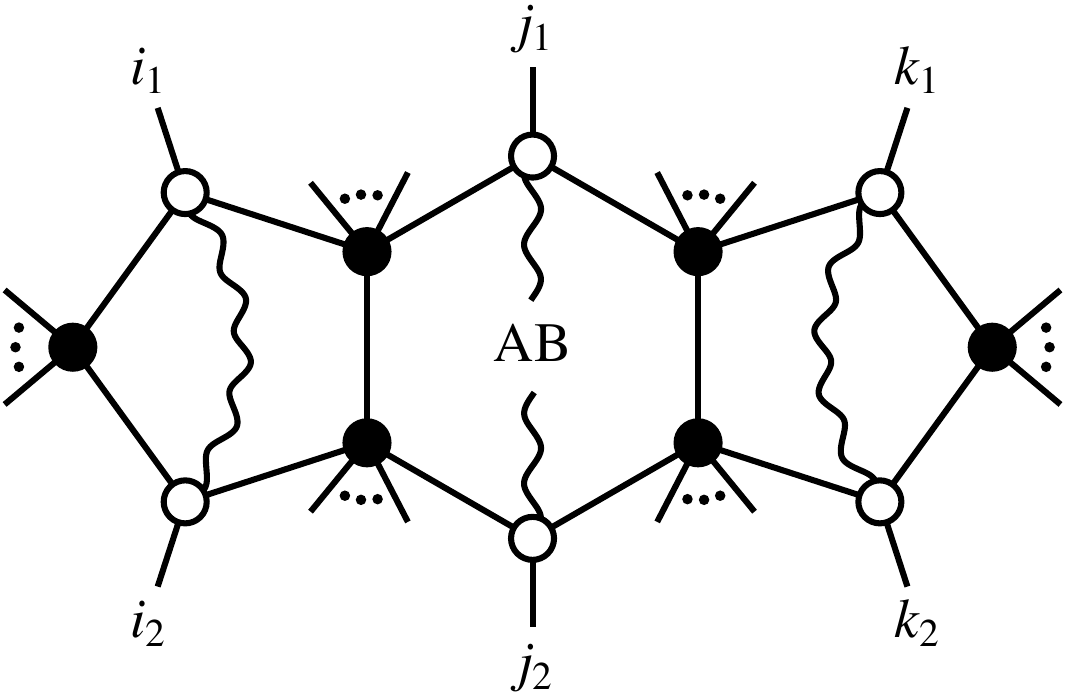}
\end{equation}
While the two-loop basis had simple boundary terms, (\ref{eq:threeloopintegrand}) hides a considerable amount of complexity that occurs at boundaries (such as $i_1 = i_2$ or $i_1=j_1$). These boundary terms are too numerous and complicated to be discussed in detail here (and can be found in the appendix of~\cite{ArkaniHamed:2010gh}), but as in the two-loop case we find that they play a crucial role in the effectiveness of the collinear, and also soft, constraints. We will therefore construct our $n$-particle three-loop basis by generating and then symmetrizing over all planar iterations of the diagrams in~(\ref{eq:threeloopintegrand}) (including the boundary terms). The goal is then to use the collinear and soft constraints to reproduce the relevant factors of 1/3 and 1/2 in the known integrand.

The size of the basis and the number of coefficients fixed by the collinear constraint are summarized through $n=11$ in table~\ref{table:threeloop}.

\begin{table}[h!]
\begin{center}
    \begin{tabular}{ c|c|c|c|}
        \cline{2-4}
    & \multicolumn{3}{|c|}{{\bf \# of unfixed coefficients}}\\
     \hline
    \multicolumn{1}{|c|}{{\bf n}} & Symmetrized basis & After collinear & After soft\\ \hline
    \multicolumn{1}{|c|}{5} & 6 & 0 & 1 \\ \hline
    \multicolumn{1}{|c|}{6} & 17 & 1 & 2 \\ \hline
    \multicolumn{1}{|c|}{7} & 33 & 2 & 0 \\ \hline
    \multicolumn{1}{|c|}{8} & 63 & 5 & 0 \\ \hline
    \multicolumn{1}{|c|}{9} & 109  & 9 & 0  \\ \hline
    \multicolumn{1}{|c|}{10} & 178  & 16 & 0  \\ \hline
    \multicolumn{1}{|c|}{11} & 277  & 26 & 0  \\ \hline
      \end{tabular}\caption{Results of the collinear and soft constraints on the three-loop basis~(\ref{eq:threeloopintegrand})}\label{table:threeloop}
      \end{center}
\end{table}

The three-loop basis is not as well-behaved as the two-loop basis under the soft limit due to the large number of boundary terms. This leads to the unfixed coefficients at five- and six-points. It is also worth noting that the collinear constraint, when applied in conjunction with the soft constraint, fixes all coefficients at $n=6$.

The collinear constraint is more effective at three-loops than at two-loops because more of the diagrams involve boxes. This points towards an interesting, though perhaps not entirely surprising feature of the collinear constraint: it grows stronger  at higher loop order. At two-loops, the efficacy of the collinear constraint decreased dramatically as $n$ increased. The data so far indicates that this is not the case at three-loops. Instead, we see that the percentage of coefficients fixed by the collinear constraint is asymptotically approaching a value somewhere near $90\%$. This is due to the large number of boundary terms involved in the three-loop basis, all of which involve boxes and are thus sufficiently divergent to be detected by the collinear limit. The total number of boundary terms at three-loops increases considerably when $n$ increases.

\section{Expanding the Basis}
The arguments of the previous section indicate that the collinear and soft constraints are useful tools for determining integrands given a suitable choice of basis. The natural question that arises, then, is what if we didn't {\it a priori} know, or at least knew very little about, the form of the integrand we wanted to construct? In other words, the chiral integrands with unit leading singularities that we used to generate bases in sections~\ref{sec:twoloop} and~\ref{sec:threeloop} are already very refined objects. If we consider a more general basis, will the collinear and soft constraints still determine the integrand?

In order to answer these questions, we will start by determining the five-point two-loop integrand from a very general basis of rational functions. We will then generalize to arbitrary $n$, using the collinear and soft constraints along with symmetry considerations to essentially re-derive the $n$-particle two-loop integrand~(\ref{eq:twoloopintegrand}).

\subsection{Five-Point Two-Loop Integrand}\label{sec:5pt2lp}
In the interest of complete generality, we will abandon diagrams and deal solely with rational functions. This means we will also have no concepts of planar or non-planar terms, which will also expand our basis considerably.

To develop a basis of rational functions, we will conjecture that they all share the common denominator
\begin{equation*}
\ab{AB12}\ab{AB23}\ab{AB34}\ab{AB45}\ab{AB51}\ab{ABCD}\ab{CD12}\ab{CD23}\ab{CD34}\ab{CD45}\ab{CD51}
\end{equation*}
We expect possible numerator terms to look like
\begin{equation}\label{eq:general} \hspace{-.5cm}
\text{\{a product of three $\ab{}$'s involving $Z_1,\ldots,Z_5$\}$\,\ab{AB\,W}\ab{AB\,X}\ab{CD\,Y}\ab{CD\,Z}$}
\end{equation}
where $W,X,Y,Z$ are bi-twistors composed out of $Z_1,\ldots,Z_5$. Taking this rational function as our seed, we iterate over all possible numerators of this form, with the only restriction being that they have the correct conformal weight.

After symmetrization, we are left with 17 linearly independent collections of these terms. Imposing the collinear and soft limits leaves us with four unfixed coefficients in addition to a constant (inhomogeneous) term. The four unfixed terms necessarily obey an interesting set of properties since they escaped detection by our methods. In particular they are all

\begin{itemize}
\item linearly independent
\item fully $D_5\times\mathfrak{s}_2$ symmetric
\item vanish under the soft limit
\item have only $\O(1/\e)$ divergences under the collinear limit.
\end{itemize}
One such object is
\begin{equation} \hspace{-.5cm}
\frac{\makebox[1cm][c]{\small $\ab{1345}\,N + \{AB \leftrightarrow CD\}+\{\mathrm{cyc.}\}$}}{\makebox[14.5cm][c]{\small $\ab{AB12}\ab{AB23}\ab{AB34}\ab{AB45}\ab{AB51}\ab{ABCD}\ab{CD12}\ab{CD23}\ab{CD34}\ab{CD45}\ab{CD51}$}}
\end{equation}
where
\begin{equation}
\begin{split}
N=-&\,\ab{2345}\ab{CD51}\ab{CD23}\ab{AB24}\ab{AB\,(123)\newcap(145)}\\
+&\,\ab{1245}\ab{CD51}\ab{CD23}\ab{AB24}\ab{AB\,(123)\newcap(345)}\\
+&\,\ab{2345}\ab{CD13}\ab{CD24}\ab{AB12}\ab{AB\,(125)\newcap(345)}\\
+&\,\ab{1345}\ab{CD12}\ab{CD24}\ab{AB25}\ab{AB\,(123)\newcap(345)}\\
+&\,\ab{1245}\ab{2345}\ab{CD23}\ab{CD13}\ab{AB12}\ab{AB45}\\
+&\,\ab{1245}\ab{2345}\ab{CD23}\ab{CD51}\ab{AB23}\ab{AB14}
\end{split}
\end{equation}
It is unclear what these objects might represent. They are clearly related in some way to non-planar diagrams, but the exact (function $\leftrightarrow$ diagram) correspondence is non-obvious.

\subsection{$n$-point Two-Loop Integrand}

For $n>5$, it is computationally prohibitive to generate a basis in the same manner as described in section~\ref{sec:5pt2lp}, so we must make some restrictions. After looking at the results for $n=4$ and 5 at two-loops, a plausible conjecture would be that two-loop integrands can be described in terms of double-pentagon integrands, i.e., rational functions with the denominator
	\vspace{.4cm}\begin{equation*}
	\makebox[1cm][c]{\small $\ab{AB\,i\smm 1\,i}\ab{AB\,i\,i\smp 1}\ab{AB\,j\smm 1\,j}\ab{AB\,j\,j\smp 1}\ab{ABCD}\ab{CD\,k\smm 1\,k}\ab{CD\,k\,k\smp 1}\ab{CD\,l\smm 1\,l}\ab{CD\,l\,l\smp 1}$}\vspace{.2cm}
	\end{equation*}
	Our task is now to determine the possible numerators for the double-pentagon integrand. In other words, we can ask: is the wavy-line numerator structure in~(\ref{eq:twoloopintegrand}) the unique numerator that satisfies the soft and collinear constraints? To answer this questions, we consider the general numerator
\begin{equation}\label{eq:conjecture}
\begin{split}
\frac{\ab{ijkl}}{\ab{ABCD}}\Big(&\frac{f(A,B,i,j)}{\ab{AB\,i\smm 1\,i}\ab{AB\,i\,i\smp 1}\ab{AB\,j\smm 1\,j}\ab{AB\,j\,j\smp 1}}\\ &\hspace{3cm}\times\frac{g(C,D,k,l)}{\ab{CD\,k\smm 1\,k}\ab{CD\,k\,k\smp 1}\ab{CD\,l\smm 1\,l}\ab{CD\,l\,l\smp 1}}\Big).
\end{split}
\end{equation}\\
Due to the symmetry of~(\ref{eq:conjecture}), we can focus our investigations on $f$:
\begin{itemize}
\item it has either odd or even parity under $\{i\leftrightarrow j\}$
\item it has odd parity under $\{A\leftrightarrow B\}$
\item it is a product of 2 $\ab{}$'s, involving one instance each of $A,B,i\smm 1,i,i\smp1,j\smm1,j,j\smp1$
\end{itemize}
The equivalent statements are true about $g$ after swapping $\{A,B,i,j\}\to\{C,D,k,l\}$. Because the product $f(A,B,i,j)g(C,D,k,l)$ must be invariant under the transformation $\{i\leftrightarrow j,k\leftrightarrow l\}$, $f$ and $g$ must have the same parity under a flip of the external indices.

To determine $f$, we start with the $70={8 \choose 4}$ possible products of $\ab{}$'s. Imposing the above symmetries leaves six linearly independent possible numerator terms that have odd parity under $\{i\leftrightarrow j\}$\footnote{One might wonder why $\ab{AB\,i\smm1\,j\smm1}\ab{i\,i\smp1\,j\,j\smp1}$ and $\ab{AB\,i\,j}\ab{i\smm1\,i\smp1\,j\smm1\,j\smp1}$ are included, but $\ab{AB\,i\smp1\,j\smp1}\ab{i\smm1\,i\,j\smm1\,j}$ is not. This is because it can be expressed as a linear combination of $o_2,o_5,$ and $o_6$.}:
\begin{equation}\label{eq:odd}
\begin{split}
f_{{\rm odd}}\,=\phantom{+}o_1 &\ab{AB\,i\,j}\ab{i\smm1\,i\smp1\,j\smm1\,j\smp1}\\
+\,o_2 &\ab{AB\,i\smm1\,j\smm1}\ab{i\,i\smp1\,j\,j\smp1}\\
+\,o_3 &\ab{AB\,(i\smm1\,i\,i\smp1)\newcap(j\smm1\,j\,j\smp1)}\\
+\,o_4 &\ab{AB\,(i\smm1\,j\,i\smp1)\newcap(j\smm1\,i\,j\smp1)}\\
+\,o_5 &\big(\ab{AB\,i\,j\smm1}\ab{i\smm1\,i\smp1\,j\,j\smp1}+\ab{AB\,i\smm1\,j}\ab{j\smm1\,j\smp1\,i\,i\smp1}\big)\\
+\,o_6 &\big(\ab{AB\,i\smp1\,j\smm1}\ab{i\smm1\,i\,j\,j\smp1}+\ab{AB\,i\smm1\,j}\ab{i\,i\smp1\,j\smm1\,j\smp1}-\\
&\phantom{\big(}\ab{AB\,i\smp1\,i\smm1\,}\ab{i\,j\smm1\,j\,j\smp1}-\ab{AB\,i\smm1\,i}\ab{i\smp1\,j\smm1\,j\,j\smp1}\big)
\end{split}
\end{equation}
There are three linearly independent possible numerator terms that have even parity:
\begin{equation}\label{eq:even}
\begin{split}\hspace{-1cm}
f_{{\rm even}}\,=\phantom{+}e_1 &\big(\ab{AB\,i\,j\smm1}\ab{i\smm1\,i\smp1\,j\,j\smp1}-\ab{AB\,i\smm1\,j}\ab{j\smm1\,j\smp1\,i\,i\smp1}\big)\\
+\,e_2 &\big(\ab{AB\, i \smp 1\,j\smm 1} \ab{i\smm1\, i\, j\, j \smp 1} -
 \ab{AB\, i \smm 1\, j} \ab{i\, i \smp 1\, j \smm 1\, j \smp 1} + \ab{AB\, i \smm 1\, j \smm 1} \times\\
& \ab{i\, i \smp 1\, j\, j \smp 1} -
 \ab{AB\, i \smm 1\, i \smp 1} \ab{i\, j \smm 1\, j\, j \smp 1} +
 \ab{AB\, i \smm 1\, i} \ab{i \smp 1\, j \smm 1\, j\, j \smp 1}\big)\\
+\,e_3 &\big(\ab{AB\, i\smp1\, j} \ab{i\smm1\, i\, j\smm1\, j\smp1} - \ab{AB\, i\, j} \ab{i\smm1\, i\smp1\, j\smm1\, j\smp1} - \ab{AB\, i\, i\smp1}\times\\
& \ab{i\smm1\, j\smm1\, j\, j\smp1} + \ab{AB\, i\smm1\, j} \ab{i\, i\smp1\, j\smm1\, j\smp1} -
 \ab{AB\, i\smm1\, i} \ab{i\smp1\, j\smm1\, j\, j\smp1}\big)\end{split}
\end{equation}
Again, the equivalent statements are true about $g$ after swapping $\{A,B,i,j\}\to\{C,D,k,l\}$.  We thus have two different possible bases generated by~(\ref{eq:conjecture}),  one involving $(f_{{\rm even}})(g_{{\rm even}})$ and the other with $(f_{{\rm odd}})(g_{{\rm odd}})$.
\subsubsection*{Even-Even Parity}
The even-even parity basis involves 6 possible $\mathfrak{s}_2$-symmetric terms generated by multiplying~(\ref{eq:even}) by the corresponding $\big(\{A,B,i,j\}\to\{C,D,k,l\}\big)$ set. Note that when constructing the basis, as outlined in section~\ref{sec:basis}, we can then absorb the $e_i$ of~(\ref{eq:even}) into the $c_i$ of~(\ref{eq:ansatz}).
Imposing the soft and collinear constraints on this basis, we find that there is no choice of coefficients able to satisfy the collinear constraint for any $n$ (checked explicitly through $n=12$). The soft constraint is possible to satisfy at $n=5$ but at no other points (again, checked explicitly through $n=12$). The even-even basis is thus not a possible basis for two-loop MHV integrands.
\subsubsection*{Odd-Odd Parity}
The odd-odd parity basis involves 21 possible $\mathfrak{s}_2$-symmetric terms generated by multiplying~(\ref{eq:odd}) by the corresponding $\big(\{A,B,i,j\}\to\{C,D,k,l\}\big)$ set. We again absorb the $o_i$ of~(\ref{eq:odd}) into the $c_i$ of~(\ref{eq:ansatz}). The size of this expanded basis and the number of coefficients fixed by the collinear constraint are summarized through $n=10$ in table~\ref{table:twoloopexp}.

\begin{table}[h]
\begin{center}
    \begin{tabular}{ c|c|c|c|}
        \cline{2-4}
    & \multicolumn{3}{|c|}{{\bf \# of unfixed coefficients}}\\
     \hline
    \multicolumn{1}{|c|}{{\bf n}} & Symmetrized basis & After collinear & After soft\\ \hline
    \multicolumn{1}{|c|}{5} & 3 & 0 & 1 \\ \hline
    \multicolumn{1}{|c|}{6} & 13 & 1 & 1 \\ \hline
     \multicolumn{1}{|c|}{7} & 33 & 3 & 0 \\ \hline
     \multicolumn{1}{|c|}{8} & 80 & 9 & 0 \\ \hline
     \multicolumn{1}{|c|}{9} & 152 & 17 & 0 \\ \hline
     \multicolumn{1}{|c|}{10} & 272 & 32 & 0 \\ \hline
      \end{tabular}\caption{Results of the collinear and soft constraints on the expanded two-loop basis~(\ref{eq:odd})}\label{table:twoloopexp}
      \end{center}
\end{table}
At $n=6$, imposing both the soft and collinear constraints fixes all coefficients.  With this data we see that the odd-odd basis, when combined with the soft and collinear constraints, is able to produce all two-loop MHV integrands. And at all $n$, the completely fixed integrand involves setting $o_3 =1$ and $o_{i\ne3}=0$ in~(\ref{eq:odd}) and then taking all planar diagrams with equal weight, reproducing the two-loop integrand~(\ref{eq:twoloopintegrand}). Furthermore, a familiar story emerges: as $n$ increases, the soft limit grows stronger while the collinear limit grows weaker. We can also compare this data with the smaller two-loop basis summarized in table~\ref{table:twoloop}. The results are intriguing: at $n=8$, for example, the collinear result constrained 64\% of coefficients with the known-numerator, but constrained 89\% of coefficients with the (much) more general numerator.

\section{Multi-Collinear Limits}
We conclude by briefly describing the behavior of integrands under multi-collinear limits, such as the double-collinear limit
\begin{equation}\label{eq:multi2}
\begin{split}
&Z_{A_1} \rightarrow Z_2 + \O(\e) ,~~~~~~~~Z_{B_1}\rightarrow  Z_1 +  Z_3 + \O(\e)\\
&Z_{A_2} \rightarrow Z_3 + \O(\e) ,~~~~~~~~Z_{B_2}\rightarrow  Z_2 +  Z_4 + \O(\e).
\end{split}
\end{equation}
As one might expect given the form of the collinear constraint~(\ref{eq:colliconst}), under this limit we have
\begin{equation}
M^{(\l)}_n\to\frac{1}{\l(\l-1)\e^4}M^{(\l-2)}_n +\O(1/\e^3).
\end{equation}
We extend this to the full $\l$-collinear limit, defined as
\begin{equation}\label{eq:multicollilim}
\begin{split}
Z_{A_1}& \rightarrow Z_{2} + \O(\e) ,\qquad Z_{B_1}\rightarrow  Z_1 +  Z_{3} + \O(\e)\\
&\vdots\hspace{4.75cm}\vdots\\
Z_{A_{\l-1}}& \rightarrow Z_{n} + \O(\e) ,\qquad Z_{B_{\l-1}}\rightarrow  Z_{n-1} +  Z_1 + \O(\e)\\
\end{split}
\end{equation}
Under this limit, we have
\begin{equation}
M^{(\l)}_n\to\frac{(-1)^\l}{\e^{2\l}\l!}+ \O(1/\e^{2\l-1}).
\end{equation}
The collinear constraint on the integrand of the logarithm can also be generalized in this manner. Specifically, $L^{(\l)}_n$ behaves as $\O(1/\e^\ell)$ in the limit~(\ref{eq:multicollilim}) for $\ell>1$. (One would na\"ively expect $\O(1/\e^{2\ell})$ divergences.)

These multi-collinear constraints are interesting from a purely conceptual standpoint, but are less useful than the single-collinear limit in terms of actually constraining a conjectured integrand basis. This is because fewer terms in the basis are sufficiently divergent to be captured by the multi-collinear constraint. This is analogous to the notion of a ``multi-soft" limit, where taking successive soft limits of external particles is essentially equivalent to removing data that would otherwise be used to construct the integrand.

\section*{Acknowledgments}
It is a pleasure to acknowledge A.~Volovich for many helpful insights.
Also we are grateful to J.~Bourjaily for his helpful insights and for the use of his figures in our paper.
This work was supported in part by US Department of Energy under contract
DE-FG02-91ER40688.

\appendix
\section{Appendix}
In this appendix we compute the collinear limit of the $n$-particle three-loop integrand, and then we show the equivalence between the collinear constraint at the level of the integrand and integrand of the logarithm for general $n$ and $\l$. Because all formulae in this appendix do not depend at all on $n$, we refer to the $n$-particle $\l$-loop integrand and integrand of the logarithm as simply $M^{(\l)}$ and $L^{(\l)}$, respectively.

Recall that $L^{(\l)}$ is given by Taylor-expanding
\begin{equation}
\log\left[1+\lambda M^{(1)}+\lambda^2 M^{(2)} +\ldots\right]
\end{equation}
to order $\l$ in $\lambda$. This Taylor expansion can be expressed recursively as
\begin{equation}
\log\left[1+\lambda M^{(1)}+\lambda^2 M^{(2)} +\ldots\right]=\sum_{\l=0}^\infty \lambda^\l L^{(\l)}
\end{equation}
where $L^{(0)}=0$, $L^{(1)} = M^{(1)}$, and for $\l>1$ we have
\begin{equation}\label{eq:basictaylor}
L^{(\l)}=M^{(\l)}-\frac{1}{\l}\sum_{i=1}^{\l-1}(\l-i)M^{(i)}L^{(\l-i)}.
\end{equation}
It is necessary to incorporate the proper symmetrization over the loop variables, so we slightly modify~(\ref{eq:basictaylor}) and arrive at the recursive definition of the (fully symmetrized) $\l$-loop integrand of the logarithm
\begin{equation}\label{eq:expansion}
L^{(\l)}=M^{(\l)}-\frac{1}{\l}\sum_{i=1}^{\l-1}(\l-i){\l \choose i}^{-1}\sum_{\text{sym}}M^{(i)}L^{(\l-i)},
\end{equation}
where
\begin{equation}\label{eq:symdef}
\sum_{\text{sym}}M^{(i)}L^{(\l-i)}\equiv \text{symmetric sum over all relevant loop variables}.
\end{equation}
We will first use this definition to examine the behavior of $M^{(3)}$ under the collinear limit, and then expand the arguments to general $\l$.

\subsubsection*{Three-Loops}
We can expand $L^{(3)}$ fully in terms of $M^{(i)}$, but it is both easier and instructive to keep some of the $L^{(2)}$ terms intact when evaluating~(\ref{eq:expansion}). This gives us
\begin{equation}\label{eq:threeloopexp}
\begin{split}
\hspace{-1cm}L^{(3)}&=M^{(3)}-\frac{1}{3}\sum_{i=1}^{2}(3-i){3 \choose i}^{-1}\sum_{\text{sym}}M^{(i)}L^{(3-i)}\\ \\
&=M^{(3)}-\frac{1}{3}\Bigg(\frac{2}{3}\left(M^{(1)}[1]L^{(2)}[2,3]+M^{(1)}[2]L^{(2)}[1,3]+M^{(1)}[3]L^{(2)}[1,2]\right)\\&\hspace{2.2cm}+\frac{1}{3}\left(M^{(2)}[1,2]L^{(1)}[3]+M^{(2)}[1,3]L^{(1)}[2]+M^{(2)}[2,3]L^{(1)}[1]\right)\Bigg).
\end{split}
\end{equation}
We have the following behavior under the collinear limit:
\begin{equation}
\begin{split}
&L^{(1)}[1]=M^{(1)}[1]\to-\frac{1}{\e^2}+\O(1/\e)\\ \\
&M^{(2)}[1,i]\to-\frac{1}{2\e^2}M^{(1)}[i]+\O(1/\e)\\ \\
&L^{(2)}[1,i]\to\O(1/\e)
\end{split}
\end{equation}
And of course $M^{(2)}[2,3]$ and $L^{(2)}[2,3]$ are unchanged under the collinear limit. With that spelled out explicitly, we take the collinear limit of~(\ref{eq:threeloopexp}):
\begin{equation}
\begin{split}
\hspace{-1.45cm}
L^{(3)}&\to\left(M^{(3)}\right)_{\rm collinear}+\frac{1}{9\e^2}\Bigg(2L^{(2)}[2,3]+M^{(1)}[2]M^{(1)}[3]+M^{(2)}[2,3]\Bigg)+\O(1/\e)\\ \\
&=\left(M^{(3)}\right)_{\rm collinear}+\frac{1}{9\e^2}\Bigg(2\left(M^{(2)}[2,3]-\frac{1}{2}M^{(1)}[2]M^{(1)}[3]\right)\\ &\hspace{5.25cm}+M^{(1)}[2]M^{(1)}[3]+M^{(2)}[2,3]\Bigg)+\O(1/\e)\\ \\
&=\left(M^{(3)}\right)_{\rm collinear}+\frac{1}{3\e^2}M^{(2)}[2,3]+\O(1/\e).
\end{split}
\end{equation}
We thus see that the three-loop integral of the logarithm vanishes at $\O(1/\e^2)$ if and only if
\begin{equation}
\left(M^{(3)}\right)_{\rm collinear}=-\frac{1}{3\e^2}M^{(2)}+\O(1/\e).
\end{equation}

\subsubsection*{$\l$-Loops}
We now wish to prove that the two definitions of the collinear constraint,
\begin{equation}
M^{(\l)} \to -\frac{1}{\l \e^2} M^{(\l-1)} +\O(1/\e)\quad(\forall\,\l>0)
\end{equation}
and
\begin{equation}
\label{eq:colliconst1}
L^{(\l)} \to \O(1/\e)\qquad\forall\,(\l>1)
\end{equation}
are equivalent at general $\l$. In order to do this, we will assume that both constraints are true through $\l-1$ and then take the collinear limit of $L^{(\l)}$ as defined in~(\ref{eq:expansion}).  Specifically, we need to take the collinear limit of the sum
\begin{equation}\label{eq:symsum}
\sum_{i=1}^{\l-1}(\l-i){\l \choose i}^{-1}\sum_{\text{sym}}M^{(i)}L^{(\l-i)}.
\end{equation}
Before we take the collinear limit, let us expand this sum out to make the relevant terms clear
\begin{equation}\label{eq:symsumexp}
\sum_{\text{sym}}M^{(i)}L^{(\l-i)} = \sum_{\text{sym'}}M^{(i)}[1,\ldots]L^{(\l-i)}[\ldots] + \sum_{\text{sym'}}M^{(i)}[\ldots]L^{(\l-i)}[1,\ldots]
\end{equation}
where the primed sum indicates the symmetric sum over the remaining loop variables $y_2,\ldots,y_\l$. In the collinear limit, we will have the following behavior
\begin{equation}
\begin{split}
&M^{(i)}[1,\ldots]\to-\frac{1}{i \e^2}M^{(i-1)}[\ldots]+\O(1/\e)\qquad \forall i>0\\ \\
&L^{(i)}[1,\ldots]\to\O(1/\e)\qquad \forall i>1\\ \\
&L^{(1)} = M^{(1)}\to-\frac{1}{\e^2}+\O(1/\e)
\end{split}
\end{equation}
Therefore the $M^{(i)}[\ldots]L^{(\l-i)}[1,\ldots]$ terms in~(\ref{eq:symsumexp}) will all vanish to $\O(1/\e^2)$ under the collinear limit, except for the $i=\l-1$ case which will become $M^{(\l-1)}$. We can now write down the $\O(1/\e^2)$ pole of~(\ref{eq:symsum}) under the collinear limit as
\begin{equation}\label{eq:expcolli1}
-\frac{1}{\l}M^{(\l-1)}- \sum_{i=1}^{\l-1}\frac{\l-i}{i}{\l \choose i}^{-1}\sum_{\text{sym}}M^{(i-1)}L^{(\l-i)}
\end{equation}
where we have dropped the prime over the symmetric sum since it is clear that the only remaining loop variables after the collinear limit are $y_2,\ldots,y_\l$. Employing a fairly straightforward juggling of summation indices of~(\ref{eq:expansion}), it is easy to show that
\begin{equation}\label{eq:expcolli2}
M^{(\l-1)} = \frac{\l}{\l-1}\sum_{i=1}^{\l-1}\frac{\l-i}{i}{\l \choose i}^{-1}\sum_{\text{sym}}M^{(i-1)}L^{(\l-i)}.
\end{equation}
Combining~(\ref{eq:expcolli1}) and~(\ref{eq:expcolli2}) tells us that under the collinear limit
\begin{equation}
\sum_{i=1}^{\l-1}(\l-i){\l \choose i}^{-1}\sum_{\text{sym}}M^{(i)}L^{(\l-i)} \to -\frac{1}{\e^2}M^{(\l-1)}+\O(1/\e).
\end{equation}
Plugging this back into our general formula for $L^{(\l)}$ gives us the following collinear behavior
\begin{equation}
L^{(\l)}\to \left(M^{(\l)}\right)_{\rm collinear} +\frac{1}{\l \e^2}M^{(\l-1)} + \O(1/\e).
\end{equation}
Therefore, under the collinear limit
\begin{equation}
M^{(\l)} \to -\frac{1}{\l \e^2} M^{(\l-1)} +\O(1/\e)\qquad \text{if and only if} \qquad L^{(\l)} \to \O(1/\e).
\end{equation}

\end{document}